\documentclass[aps,prb,twocolumn,superscriptaddress,amsmath,amssymb,notitlepage]{revtex4-2}
\usepackage{graphicx}
\usepackage{amsmath}
\usepackage{amssymb}
\usepackage{float}
\usepackage{color}

\usepackage{cancel,xcolor}
\usepackage{indentfirst}
\usepackage{CJKulem}
\usepackage{soul,xcolor}
\setstcolor{red}

\usepackage{ulem}
\usepackage{cancel}
\usepackage[pdfpagemode=UseNone,pdfstartview=FitH,colorlinks=true,linkcolor=blue,urlcolor=blue,anchorcolor=blue,citecolor=blue]{hyperref}

\newcommand{\beq}{\begin{equation}}
\newcommand{\eeq}{\end{equation}}
\newcommand{\beqa}{\begin{eqnarray}}
\newcommand{\eeqa}{\end{eqnarray}}

\def\Aa2#1{\textcolor{magenta}{#1}}

\def\Aa1#1{\textcolor{blue}{#1}}
\def\prb#1{{ Phys.\ Rev. B\/} {\bf#1}}

\def\prl#1{{ Phys.\ Rev.\ Lett.} {\bf#1}}

\begin{document}

\title{Different critical behaviors in cubic to trigonal and tetragonal perovskites}

\author{Amnon Aharony}
\email{aaharonyaa@gmail.com}
\affiliation{ School of Physics and Astronomy, Tel Aviv University, Tel Aviv 6997801, Israel}

\author{Ora Entin-Wohlman}
\email{orawohlman@gmail.com}
\affiliation{ School of Physics and Astronomy, Tel Aviv University, Tel Aviv 6997801, Israel}

 \author{Andrey Kudlis}
\email{andrewkudlis@gmail.com}
\affiliation{ITMO University, Kronverkskiy prospekt 49, Saint Petersburg 197101, Russia}

\begin{abstract}

Perovskites like LaAlO$^{}_3$  (or SrTiO$^{}_3$) undergo displacive structural phase transitions from a cubic crystal to a trigonal (or tetragonal) structure. For many years, the critical exponents in both these types of transitions have been fitted  to those of the isotropic three-components Heisenberg model.  However,  field theoretical calculations showed that the isotropic fixed point of the renormalization group is unstable, and renormalization group iterations flow either to a cubic fixed point or to a fluctuation-driven first-order transition.
Here we show that these two scenarios correspond to the cubic to trigonal and to the cubic to tetragonal transitions, respectively. In both cases, the critical behvior is described by slowly varying effective critical exponents, which exhibit universal features. 
 For the trigonal case, we predict a crossover of the effective exponents from their Ising values to their cubic values (which are close to the isotropic ones). For the tetragonal case, the effective exponents can have the isotropic values over a wide temperature range, before exhibiting large changes en route to the first-order transition. New renormalization group calculations near the isotropic fixed point in three dimensions are presented and used to estimate the effective exponents, and dedicated experiments to test these predictions are proposed. Similar predictions apply to cubic magnetic and ferroelectric systems.

\end{abstract}

%%%%%%%%%%%%%%%%%%%%%%%%%%%%%%%%%%%%%%%%%%%%%%%%%
%%%%%%%%%%%%%%%%%%%%%%%%%%%%%%%%%%%%%%%%%%%%%%%%%

\date{\today}
\maketitle
%%%%%%%%%%%%%%%%%%%%%%%%%%%%%%%%%%%%%%%%%%%%%%%%%
%%%%%%%%%%%%%%%%%%%%%%%%%%%%%%%%%%%%%%%%%%%%%%%%%

\section{Introduction}
\label{Intro}

\begin{figure}
\centering
\includegraphics[width=0.2\textwidth]{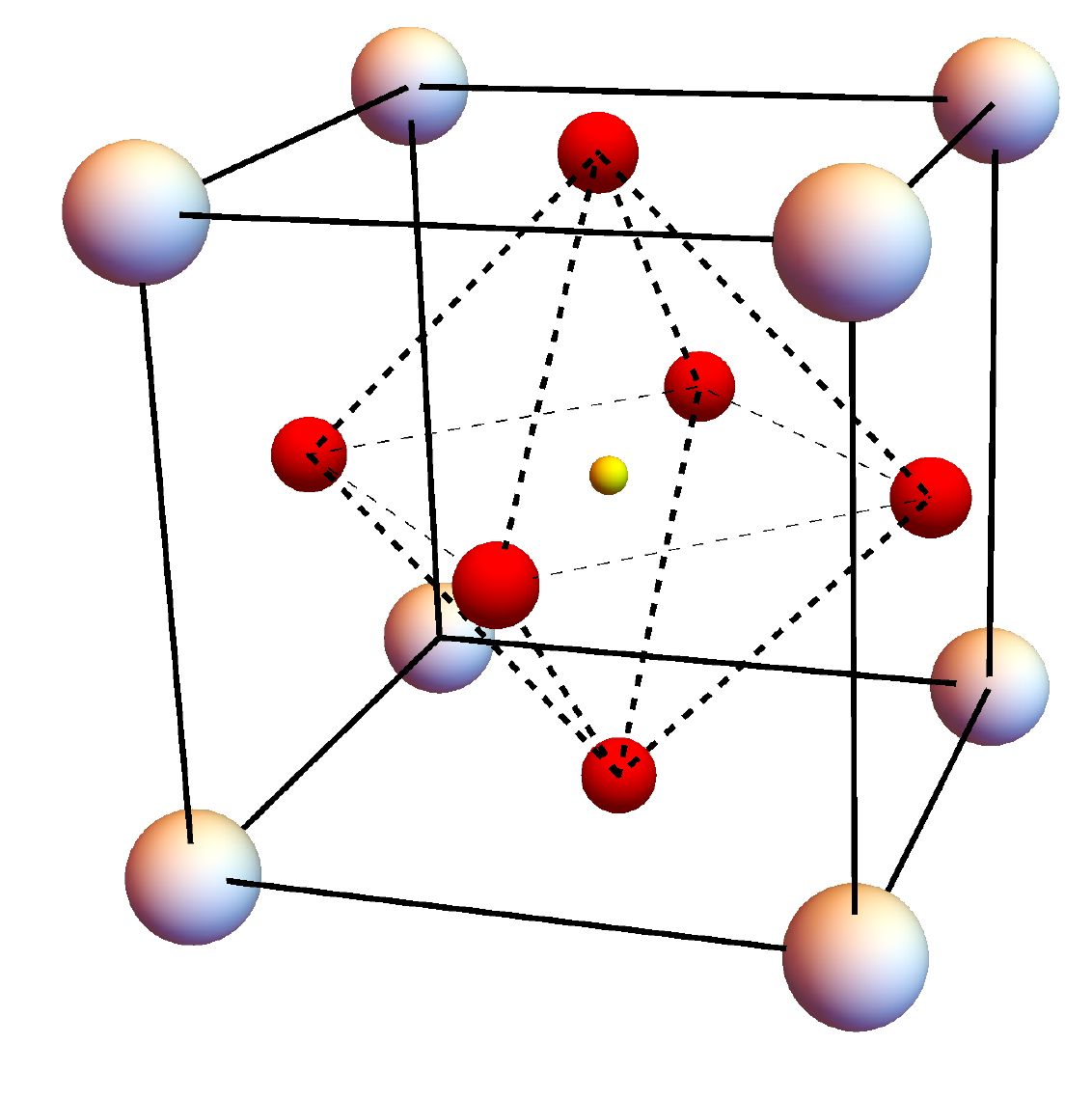}\ \ \ \ \ \includegraphics[width=0.2\textwidth]{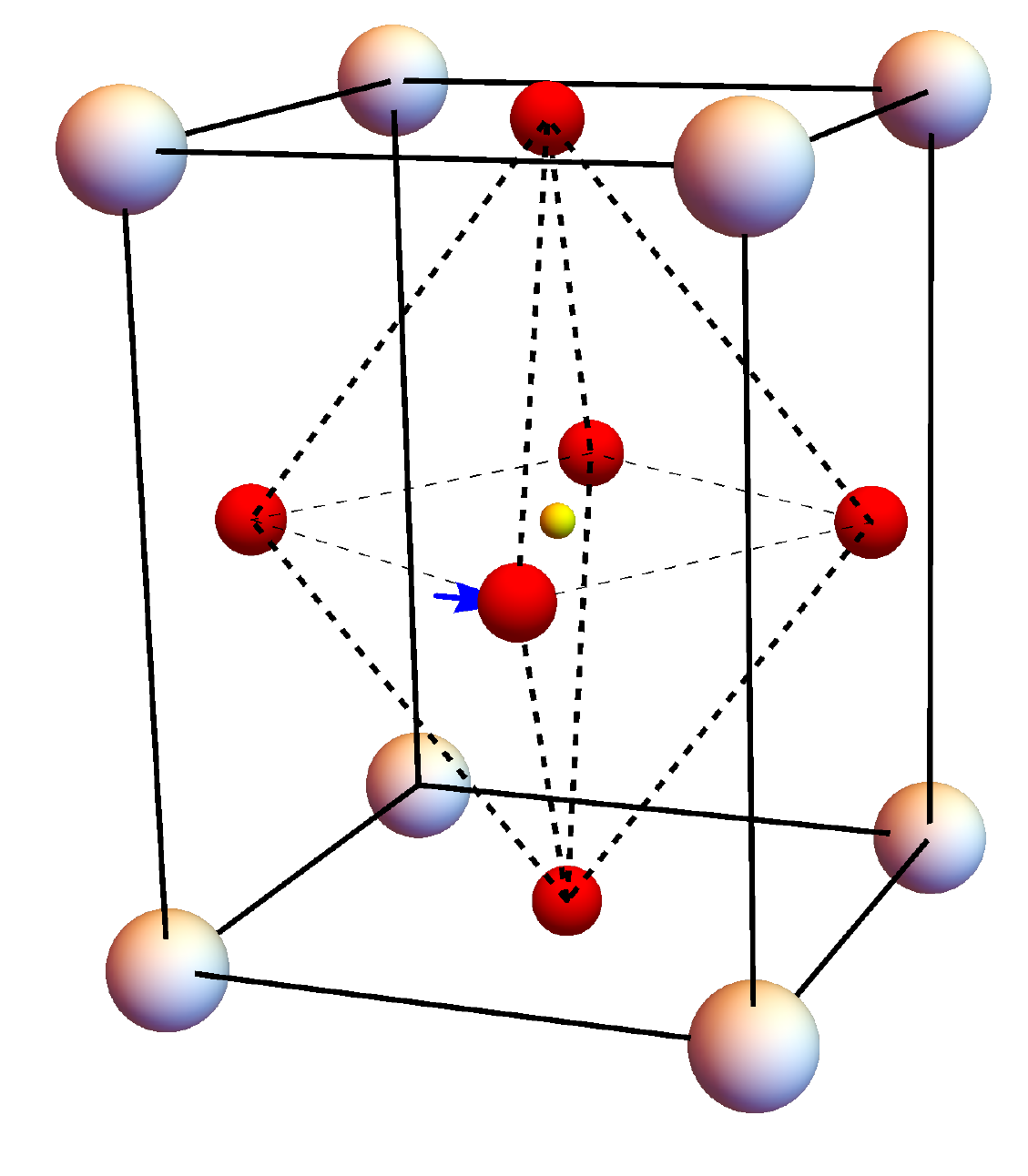}
\caption{(color online) The cubic (left) and tetragonal (right) unit cells in SrTiO$_3$ (the latter shows only half the cell: neighboring cells rotate in opposite directions). Large, intermediate and small spheres correspond to Sr, O and Ti ions, respectively. The dashed lines represent the octahedra, which rotate around the vector ${\bf Q}$, lying along the vertical axis (the O ions in the central horizontal plane move as indicated by the arrow). When ${\bf Q}$ is along a diagonal of the cube, the octahedra rotate around that  diagonal, and the  unit cell  is stretched along ${\bf Q}$, causing a cubic to trigonal transition (as in LaAlO$^{}_3$).}
\label{Fig1}
\end{figure}

%\noindent\textit{The perovskites.--}
Perovskite materials exhibit intriguing physical properties, and have been extensively explored for both practical applications and theoretical modeling~\cite{perov}. In particular, perovskites like SrTiO$^{}_3$ and LaAlO$^{}_3$ play important roles in modern solid state applications~\cite{STO-LAO}. At high temperatures,  perovskites usually have a cubic structure (left panel, Fig. \ref{Fig1}).  As the temperature $T$ decreases, some perovskites undergo an antiferrodistortive structural transition from the cubic to a  lower-symmetry structure, via a rotation of the oxygen (or fluorine) octahedra:  SrTiO$^{}_3$, KMnF$^{}_3$, RbCaF$^{}_3$  and  others undergo a cubic to tetragonal transition, see Fig. \ref{Fig1}.  The octahedra rotate around a cubic axis and the order-parameter vector
 ${\bf Q}$ (aka the rotation vector) is along that axis (with a length proportional to the rotation angle).
 Similar rotations occur in  double perovskites, e.g., the tetragonal to orthorhombic transition in La$^{}_2$CuO$^{}_4$~~\cite{axe}.  In contrast, other perovskites, e.g., LaAlO${}_3$, PrAlO$^{}_3$, and  NdAlO$^{}_3$, undergo a cubic to trigonal transition, with ${\bf Q}$ along one of the cubic diagonals.

%%%%%%%%%%%%%%%%%%%%%%%%%%%%%%%%%%%%%%%%%%%%%%%%%%%%%%%%%%%%%%%%%%%%%
%%%%%%%%%%%%%%%%%%%%%%%%%%%%%%%%%%%%%%%%%%%%%%%%%%%%%%%%%%%%%%%%%%%%%

 The  behavior of a system at the vicinity of its transition temperature $T^{}_{c}$ can be expressed by  critical exponents. When the transition at $T^{}_c$  is continuous, the correlation length diverges as $\xi\propto |t|^{-\nu}$ and the order-parameter approaches zero (for temperatures $T<T^{}_c$) as $|\langle{\bf Q}\rangle|\propto |t|^\beta$, where $t=T/T^{}_c-1$. The critical exponents $\nu$ and $\beta$ are expected to be universal, i.e., having the same values for many physical systems. The exponents describing other physical properties, e.g., $\alpha$ and $\gamma$ for the specific heat and for the order parameter susceptibility, are obtained via scaling relations, $d\nu=2-\alpha=2\beta+\gamma$, where $d$ is the dimensionality.

The $T-$dependence of the order-parameters of SrTiO$^{}_3$ and LaAlO$^{}_3$ had been measured by  M\"{u}ller and Berlinger~\cite{KAM1970}, who obtained a  collapse of the scaled order-parameters in both materials onto a single line, showing a crossover from the  mean-field exponent $\beta^{}_{MF}=1/2$, valid relatively far away from $T_{c}$ (i.e., at relatively large $|t|$) to an apparently universal critical exponent $\beta^{}_c=0.33\pm0.02$ close to $T^{}_c$.
Consequently, it was  concluded erroneously that both materials  share the same exponents, and so belong to the {\it same universality class} and have the same critical behavior. In fact it  turned out that both samples were {\it single domains}, due to a uniaxial anisotropy caused by polishing~\cite{AB1}. Therefore both systems ordered along a {\it single} axis, and the data were consistent with Ising ($n=1$, a single-component order-parameter) exponents~\cite{KAM2,AB2,paris}. Later experiments gave a wide range of results, e.g.  $\beta\simeq 0.40\pm 0.03,~1/3,~0.27,~0.17\pm 0.02$
for SrTiO$^{}_3$~~\cite{KAM-F}, KMnF$^{}_3$~~\cite{borsa}, RbCaF$^{}_3$ and NaNbO$^{}_3$~~\cite{NaNbO3}, respectively. Furthermore, some experiments hint that SrTiO$^{}_3$ may be close to a tricritical point ~\cite{garnier,STO1st,Gallardo}, while both RbCaF$^{}_3$ and KMnF$^{}_3$ have  first-order transitions~\cite{Rb,stokka}. More
 experiments are reviewed e.g. in Refs. \onlinecite{NATO,KAMrev,cowley}.

%%%%%%%%%%%%%%%%%%%%%%%%%%%%%%%%%%%%%%%%%%%%%%%%%%%%%%%%%%%%%%%%%%%%%
%%%%%%%%%%%%%%%%%%%%%%%%%%%%%%%%%%%%%%%%%%%%%%%%%%%%%%%%%%%%%%%%%%%%%

Here we show theoretically that the cubic to trigonal and the cubic to tetragonal phase transitions do not share the same critical behavior. In fact these two transitions are characterized by different behaviors of the renormalization-group (RG) iterations.  We find that systems undergoing a cubic to trigonal transitions  have second-order transitions, with the universal critical exponents of the cubic fixed point (FP). However, since the cubic and the isotropic fixed points turn out to be very close to each other (see below), it is indeed difficult to distinguish between their asymptotic critical exponents even without any uniaxial symmetry breaking.  As we show, these asymptotic exponents are expected only very close to $T^{}_c$. For a large range of $t$ we predict {\it effective exponents}, which vary with $t$. In contrast, the cubic to tetragonal transitions become fluctuation-driven first-order at small $|t|$, with different effective exponents over an intermediate range of $t$.  These effective exponents are characterized by the isotropic fixed point, which dominates the crossover to the first-order transition. This conclusion extends  to quite a number of other systems as well. %The results are derived within the theory of (finite-temperature) phase transitions and critical behavior.

Our predictions for the effective exponents come from  a novel way to obtain analytic solutions to the RG recursion relations in the vicinity of the isotropic and cubic fixed points.
After a short review of the theory of the RG on cubic systems in Sec. II, Sec. III presents our new calculation.
Our predictions are compared with the above experimental information in Sec. IV, where we also  propose dedicated measurements to test our results and list other cubic systems, which are expected to exhibit similar behaviors. Section V contains our conclusions. The Appendices describe the resummation technique and the analytic solution of the recursion relations.

%%%%%%%%%%%%%%%%%%%%%%%%%%%%%%%%%%%%%%%%%%%%%%%%%%%%%%%%%%%%%%%%%%%%%
%%%%%%%%%%%%%%%%%%%%%%%%%%%%%%%%%%%%%%%%%%%%%%%%%%%%%%%%%%%%%%%%%%%%%

 \section{ The renormalization group on cubic systems}

%%%%%%%%%%%%%%%%%%%%%%%%%%%%%%%%%%%%%%%%%%%%%%%%%
%%%%%%%%%%%%%%%%%%%%%%%%%%%%%%%%%%%%%%%%%%%%%%%%%

 Fifty years ago, Wilson \cite{wilson} showed that at $T$ very close to $T^{}_c$  the short-length details can be eliminated  on scales below $1/e^\ell$ ($\ell$ counts the number of iterations in the elimination process),
 and that rescaling the unit length by the factor $e^\ell$ yields a renormalized effective (dimensionless) Hamiltonian (or free-energy density) $\overline{\cal H}(\ell)$, which `flows' in the space spanned by all such Hamiltonians. These flows represent the RG.
 Critical points are associated with fixed points of these flows, which are invariant under the RG iterations.  Near a fixed point,
 the singular part of the
 corresponding free-energy density  obeys the homogeneous scaling form
\begin{align}
{\cal F}(\{\mu^{}_i\})&=e^{-d\ell^{}_f}{\cal F}(\{\mu^{}_i(0)e^{\lambda^{}_i\ell^{}_f}\})\nonumber\\
&\equiv|t|^{d\nu}{\cal W}(h|t|^{-\nu\lambda^{}_2},~\mu^{}_3(0)|t|^{-\nu\lambda^{}_3},~\dots),
\label{FW}
\end{align}
 where the $\mu^{}_i(0)$'s are parameters that measure deviations from the fixed point, and the $\lambda^{}_i$ are exponents describing their variation as function of $\ell$. The first two parameters are the temperature $\mu^{}_1=t$ and the ordering field $\mu^{}_2=h$. Continuing the RG flow until  $\xi(\ell^{}_f)=\xi(0)e^{-\ell^{}_f}\sim 1$, with $\xi(0)\sim |t(0)|^{-\nu}$, yields  $\nu=1/\lambda^{}_1$.
 Derivatives of ${\cal F}$ w.r.t. $t$ and $h$ yield the critical exponents for the measurable quantities, e.g. the scaling relations $\beta=\nu(d-\lambda^{}_2)$, $\gamma=\nu(d-2\lambda^{}_2)$ and $\alpha=2-d\nu$.
 The exponents and the scaling function ${\cal W}$ for systems near a specific FP are fully determined by the FP itself, and not by the initial effective Hamiltonian,  which encompasses the short-scales behavior. Therefore they are universal~\cite{WK,MEF1,MEF,DG}.  All the physical systems which flow to that FP then belong to its universality class, and exhibit the same critical exponents. A stable fixed point has only two relevant variables, $t$ and $h$, with $\lambda^{}_1,~\lambda^{}_2>0$. All the  other parameters are irrelevant, with $\lambda^{}_i<0$. In contrast, when a third parameter, $\mu^{}_3$, is also relevant, $\lambda^{}_3>0$, the fixed point is unstable, and the RG trajectories flow to another, stable, fixed point, via a crossover region, or flow to a region where the renormalized Hamiltonian has a first order transition.

 The RG analysis of cubic systems has been mainly based on the Landau theory~\cite{landau,toledano},  which expands
  the free-energy density in powers of the (small) order-parameter components $Q^{}_i$  ($i=1,2,\ldots,n$). The terms in this expansion are determined by  the symmetries of the system {\it above the transition}.
 For the isotropic $n-$component order-parameter vector ${\bf Q}$, this free-energy is
 $U^{}_0({\bf Q})=r|{\bf Q}|^2/2+u|{\bf Q}|^4+{\cal O}[|{\bf Q}|^6]$, where  $u$ is a system-dependent parameter, while $r=T/T^{\rm MF}_c-1$, with the mean-field transition temperature $T^{\rm MF}_c$ (the parameter  $t$, mentioned above, contains the downwards shift from $T^{\rm MF}_c$ to  $T^{}_c$ by the fluctuations).
  The cubic symmetry is characterized by adding the term $U^{}_v({\bf Q})=v\Sigma^n_{i=1} Q^4_i$  to the free energy, with $v$ a system-dependent coefficient~~\cite{ST,AA1973}. Long wave-length fluctuations in ${\bf Q}({\bf r})$  are introduced via a  gradient term~\cite{GL}, $|{\boldmath{\nabla}}{\bf Q}({\bf r})|^2$,
whose coefficient is normalized to $1$.
The  effective Hamiltonian is then written as $\int d^dr \overline{\cal H}({\bf r})$,  where
\begin{align}
\overline{\cal H}({\bf r})\equiv|{\boldmath{\nabla}}{\bf Q}({\bf r})|^2/2+U^{}_0[{\bf Q}({\bf r})]+U^{}_v[{\bf Q}({\bf r})].
\label{H0}
\end{align}
 In the absence of the ordering field $h$, the parameters which flow under the RG iterations are $r~,u$ and $v$.

The Wilson-Fisher RG  at $d=4-\epsilon$ is performed in Fourier space, eliminating large momentum (small length scale) components of the order-parameter~\cite{WF,WK}. As we discuss below, it is not trivial to extrapolate the results to $d=3$, i.e., $\epsilon=1$.
Generally, the recursion relations in the $u-v$ plane have the form
\begin{align}
\frac{\partial u}{\partial \ell}&=\bar{\beta}^{}_u[\epsilon,u(\ell),v(\ell)],\ \ \ \frac{\partial v}{\partial \ell}=\bar{\beta}^{}_v[\epsilon,u(\ell),v(\ell)],
\label{RR}
\end{align}
and the $\bar{\beta}$ functions (not to be confused with the critical exponent $\beta$) are expanded in powers of their arguments. The fixed points $u^\ast,v^\ast$ are found as the zeroes of these functions, with values which are series in $\epsilon$.
For $v=0$, this procedure
gives two FP's, one at $u^\ast_G=0$, termed Gaussian, and the  other, termed the  isotropic FP, with $u^\ast_I(n)={\cal O}[\epsilon]>0$. For $d<4$ the Gaussian FP is unstable, i.e., $\lambda^G_u>0$, so that systems for which $u>0$  flow towards the stable isotropic FP~\cite{Xover} (where $\lambda^I_u<0$) and those with $u<0$ flow to a region in which the mean-field analysis of the renormalized free-energy yields  a first-order transition (stabilized by the positive sixth-order terms)~\cite{1stordr,blank}. The Gaussian FP is thus identified as a tricritical point, separating between a first- and a second-order transitions.

%%%%%%%%%%%%%%%%%%%%%%%%%%%%%%%%%%%%%%%%%%%%%%%%%%%%%%%%%%%%%%%%%
%%%%%%%%%%%%%%%%%%%%%%%%%%%%%%%%%%%%%%%%%%%%%%%%%%%%%%%%%%%%%%%%%

\begin{figure}[htb]
\vspace{-1.4cm}
\includegraphics[width=0.56\textwidth]{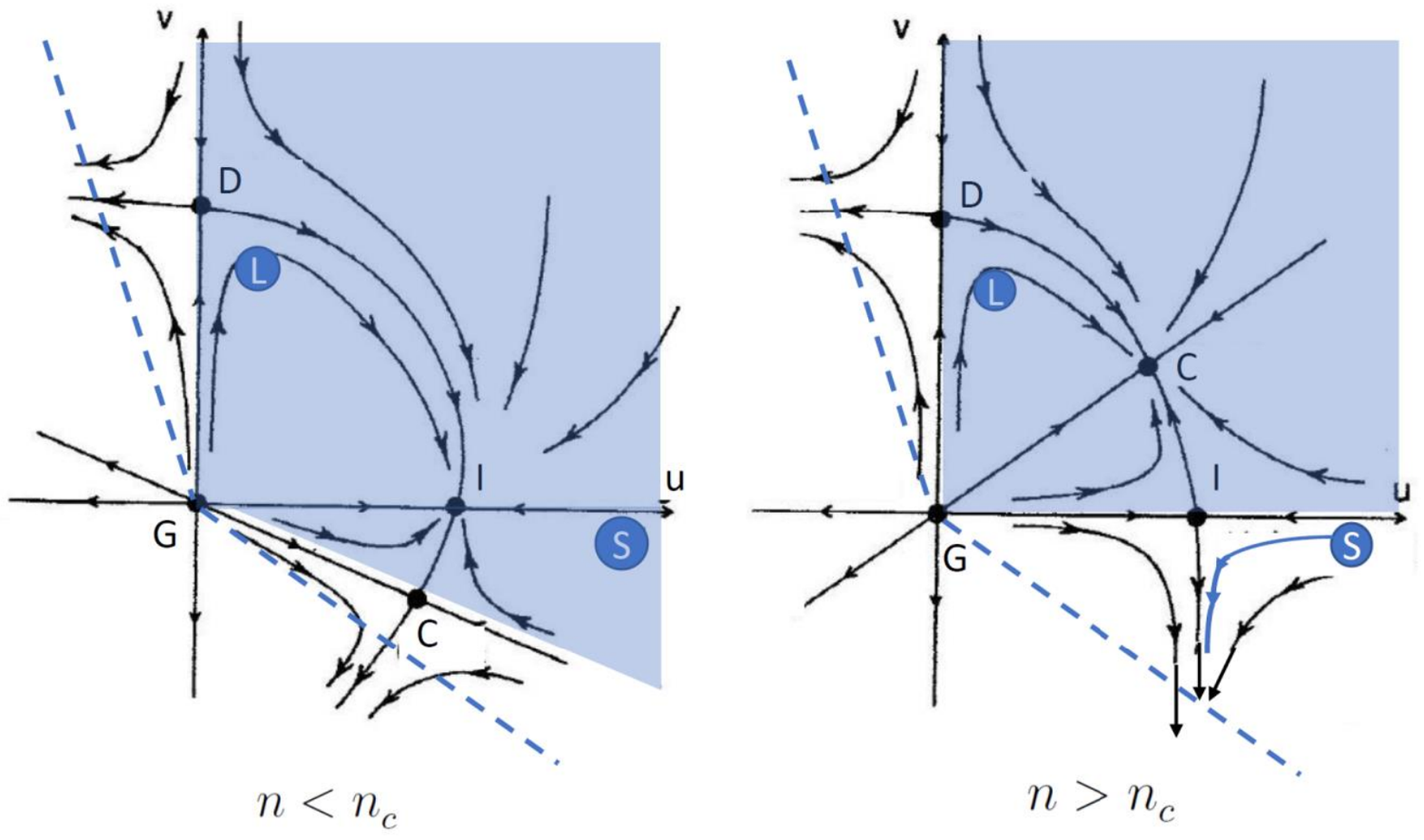}
\vspace{-8.7cm}
\caption{(color online)  Schematic flow diagram and fixed points for the cubic model, Eq. (\ref{H0}), adapted from Ref. \onlinecite{DG}. %Left: $n<n^{}_c$. Right: $n<n^{}_c$.
$G=$Gaussian, $I=$isotropic, $D=$Decoupled (Ising) and $C=$Cubic FP's. S=initial point for SrTiO$^{}_3$. L=initial point for LaAlO$^{}_3$. The dashed lines represent the stability edges, $u+v=0$ (for $v<0$) and $u+v/n=0$ (for $v>0$), below which  the free energy in Eq. (2) is stabilized by the terms of order $|{\bf Q}|^6$, and the transitions are first-order. The shaded areas are the regions of attraction of the stable FP's ($I$ on left and $C$ on right). }
\label{fp}
\end{figure}

The RG analysis  of the cubic model, Eq. (\ref{H0}), for general $n$ %(the case $n=2$ was included in Ref. \onlinecite{WF})   %
and in  dimension $d=4-\epsilon$ ~~\cite{AA1973,BC1973,Wallace1973} yielded four FP's of order $\epsilon$:   the Gaussian ($G$, $u^\ast_G=v^\ast_G=0$),  isotropic ($I$, $v^{\ast}_I=0,~u^\ast_I>0$), decoupled Ising ($D$, $u^{\ast}_D=0,~v^\ast_D>0$,  for which the different $Q^{}_i$'s decouple from each other and exhibit  the Ising model behavior),  and  `cubic' ($C$) FP's. The location of the cubic FP, $(u^{\ast}_C,~v^{\ast}_C)$,  depends on the number of the order-parameter components,
 $n$: for small (large) $n$, it is in the lower (upper) half plane, as shown on the  left (right) panel of Fig. \ref{fp}.
 %, and only the isotropic (cubic) FP is stable.see  Fig. \ref{fp}.
 This figure~\cite{DG} has been  reproduced by  other authors, e.g., Refs. \onlinecite{6loops,eps6,vic-rev}, and included in textbooks~\cite{CL}. In these references, the FP values were calculated using various approximations, but (except at lowest order in $\epsilon$) the flow lines were drawn schematically.  The figure shows the critical surface of the effective Hamiltonian in the $u-v$ plane, on which $|t|=h=0$ and $\xi=\infty$~~\cite{FSS}. At a finite (but very small) $|t|$ the RG flow starts very close  to this surface, and as the RG is iterated the flow stays close to the arrows in the figure. When the flow reaches the vicinity of a FP %(after $\ell^{}_1$ iterations),
 the system exhibits  the critical exponents  of that FP.

%%%%%%%%%%%%%%%%%%%%%%%%%%%%%%%%%%%%%%%%%%%%%%%%%
%%%%%%%%%%%%%%%%%%%%%%%%%%%%%%%%%%%%%%%%%%%%%%%%%

%%%%%%%%%%%%%%%%%%%%%%%%%%%%%%%%%%%%%%%%%%%%%%%%%
%%%%%%%%%%%%%%%%%%%%%%%%%%%%%%%%%%%%%%%%%%%%%%%%%

%\noindent\textit{Competing fixed points.--}
As seen in Fig. \ref{fp}, the Gaussian FP is doubly unstable; both $u$ and $v$ are relevant in its vicinity. The decoupled FP is singly unstable, with $u$  being relevant~\cite{DG,AA}. %with a `stability exponent' equal to the specific heat exponent of the Ising model, $\lambda^D_v=\alpha^{}=2-d\nu^{}$, equal to the exponent of the Ising model specific heat~\cite{DG,AA}.
The stability of the isotropic and cubic FP's depends on the borderline value $n^{}_c(3)$.
It was clear that $2<n^{}_c(3)<4$, but  different approximations
yielded conflicting answers  to the question whether $n=d=3$ is above or below $n^{}_c(3)$ (for the history, see Refs. \onlinecite{paris,MC,6loops,eps6,boot,vic-rev}  and references therein).
For instance, a third-order $\epsilon-$expansion gave \cite{DG} $n^{}_c(3)\approx 3.128$ at $\epsilon=1$. The result
$3<n^{}_c(d=3)$ was also obtained by the scaling-field method~~\cite{DOMR}.
If this were true then all the second order transitions cubic systems would be decsribed by the universality class of the isotropic FP, and the cubic deviatoions from rotational symmetry would decay at criticality.

However, this scenario is now known to be wrong. 
 Four  accurate  methods (Monte Carlo simulations of lattice $O(n)$ models~\cite{MC}, six-loop recursion relations~\cite{6loops} at $d=3$, the good old $\epsilon$-expansion,  recently expanded to order $\epsilon^6$~~\cite{eps6} and the very recent bootstrap method, which calculates exponents at any dimension~\cite{boot}) find  $2.85<n^{}_c(3)<3$.
 Therefore, the RG flows are as in the right panel in Fig. \ref{fp}: the isotropic FP is {\it unstable} [with a small but positive exponent for the flow of $v$, $0<\lambda^I_v\backsimeq 0.02$, see Eq. (1)],  while the cubic  FP has a small but positive FP value $v^{\ast}_C>0$, and is fully stable (deviations of both $u$ and $v$ from it decay under  the RG iterations).

Without even looking at the specific numerical values of the locations of the FP's, the right  panel of Fig. \ref{fp} yields {\it qualitative} crucial consequences: since the cubic to trigonal and cubic to tetragonal transitions correspond to opposite signs of $v$, they have different flow trajectories.  For the former, $v<0$. If
  the initial $|v|$ is small, and if the initial $u$ is positive, so that $u+v>0$, then the respective  effective Hamiltonian (shown by the blue trajectory leaving S in Fig. \ref{fp}) first flows closer to the isotropic FP, and may then exhibit $t-$dependent effective exponents associated with that FP, but eventually it {\it must} turn downwards, and cross the stability line $u+v=0$ - turning the transition fluctuation-driven first-order~\cite{dom3}.  As $\lambda^I_v$  is  small at $n=d=3$~~\cite{MC,6loops,eps6,boot}, this flow is   slow, so that the first-order transition will occur only  close to $T_{c}$, with a small discontinuity. For larger initial $|v|$'s the transition becomes first-order at larger $|t|$. In contrast, the cubic to trigonal transition has $v>0$, and therefore its  Hamiltonian  {\it must} flow to the stable cubic FP, resulting in a second-order transition with cubic exponents.

 As mentioned above,  the cubic and the isotropic FP's are close to each other at $n=d=3$~~\cite{MC,6loops,eps6,boot}. Indeed, the calculated asymptotic critical exponents (expected only at very small $|t|$) are $\nu^{I}\backsimeq 0.706,~\nu^C\backsimeq 0.700$, $\beta^{I}_{}\backsimeq .366,~\beta^{C}_{}\backsimeq0.368$. This closeness also implies that the stability exponent of the stable cubic FP, $\lambda^C_3\lesssim 0$ %(associated with  $\mu^{}_3(0)$, which is a combination of $v$ and $u$ having this small exponent)
 and that of the unstable isotropic FP, $\lambda^I_v\gtrsim 0$, are small, indicating slow flows towards and away from these FP's. The experimental exponents should therefore be compared with effective exponents, e.g., $\beta^{}_{\rm eff}\equiv \partial \log |{\bf Q}|/\partial\log|t|$, which depend on $|t|$.
 For $v>0$, the cubic FP is stable, with two negative stability exponents, $\lambda^C_4<\lambda^{C}_3<0$ (for flows in the $u-v$ plane). Therefore
 $\beta^{}_{\rm eff}$ approaches
  the  asymptotic $\beta^{C}_{}$ slowly, with corrections of order $|t|^{|\varphi^C_3|}$, where $\varphi^C_3=\nu^C\lambda^C_3$ is small.
  For $v<0$, $\beta^{}_{\rm eff}$ first approaches the isotropic FP value $\beta^{I}_{}$, but then (for smaller $|t|$, i.e., larger number of iterations $\ell$), it moves away from that value, and $|{\bf Q}|$ has a
   discontinuity.

%%%%%%%%%%%%%%%%%%%%%%%%%%%%%%%%%%%%%%%%%%%%%%%%%%%%%%%
%%%%%%%%%%%%%%%%%%%%%%%%%%%%%%%%%%%%%%%%%%%%%%%%%%%%%%%

\section{RG flow near the isotropic and cubic fixed points}

%%%%%%%%%%%%%%%%%%%%%%%%%%%%%%%%%%%%%%%%%%%%%%%%%%%%%%%
%%%%%%%%%%%%%%%%%%%%%%%%%%%%%%%%%%%%%%%%%%%%%%%%%%%%%%%

 To quantify the above qualitative statements, we used existing  $\epsilon^6-$order expansions~\cite{eps6} to derive RG flow equations in the vicinity of the isotropic and cubic fixed points at $d=n=3$.
As discussed in Ref. \onlinecite{eps6} (and  also in other references, e.g., Refs.~\onlinecite{brezin1,LIPATOV,KP17,KW20}), these series are  divergent, and numerical estimates of the quantities of interest  at $\epsilon=1$ were obtained employing  resummation techniques, see Appendix  \ref{app:resummation_strat}.

Critical exponents are found from the RG recursion relations for the various scaling fields. To linear order in $\mu^{}_i(\ell)$ these  have the form
\begin{align}
\frac{\partial \mu^{}_i}{\partial \ell}=\lambda^{}_i(\epsilon,u,v)\mu^{}_i(\ell),
\label{mu}
\end{align}
 where the effective exponent $\lambda^{}_i$ is expanded in powers of its arguments and then resummed. At a fixed point, $u^\ast$ and $v^\ast$ are replaced by their  $\epsilon-$expansions,  and the series for $\lambda^{}_i$ are resummed to give asymptotic values at $d=n=3$~\cite{eps6}.

As mentioned,  $v$ varies slowly near the isotropic and cubic FP's. Therefore, we need to solve the recursion equations (\ref{RR}) as functions of the number of iterations $\ell$, and then estimate the effective critical exponents $\lambda^{}_i(\ell)$ at finite values of $\ell$, up to the total number of iterations $\ell^{}_f$. Since $t(\ell)=t(0)e^{ \lambda^{}_1\ell}$, a larger $\ell^{}_f$ implies a smaller $t(0)$, i.e., an initial state closer to the critical point.
 The divergence of the  $u-$ and $v-$dependent series for $\beta^{}_u$, $\beta^{}_v$ and $\lambda^{}_i$ prevents their use in solving the differential equations (\ref{RR}). Instead, we derived approximate expressions of $\beta^{}_u$ and $\beta^{}_v$, by expanding them  to second order near the isotropic FP, and resumming the $\epsilon-$expansions of the resulting coefficients. Since the cubic FP is expected to be close to the isotropic one, we expect these expansions to give reasonable results near both FP's. Hence,
\begin{align}
&\frac{\partial \delta u}{\partial\ell}=\lambda^{}_u\delta u+a^{}_{01}v+a^{}_{11}v \delta u
+\frac{1}{2}[a^{}_{20}\delta u^2+a^{}_{02}v^2],
\label{RG1}\\
&\frac{\partial v}{\partial\ell}=\lambda^{}_v v+ b^{}_{11}v \delta u
+\frac{1}{2}b^{}_{02}v^2,
\label{RG2}
\end{align}
where $\delta u=u-u^\ast_I$ and the coefficients $a^{}_{ij}$ and $b_{ij}$,  which are given in  Table \ref{one},
 are found from
\begin{eqnarray}
    &&\left.a^{}_{ij}=\widehat{\textup{Rsm}}\Bigg[\dfrac{\partial^{(i+j)}\beta_{u}(u,v)}{\partial u^i \partial v^j}\Bigg]\right\rvert_{u=u^\ast_I,v=0},\nonumber\\         &&\left.b^{}_{ij}=\widehat{\textup{Rsm}}\Bigg[\dfrac{\partial^{(i+j)}\beta_{v}(u,v)}{\partial u^i \partial v^j}\Bigg]\right\rvert_{u=u^\ast_I,v=0}.
\end{eqnarray}
The operation $\widehat{\textup{Rsm}}$ denotes  resummation of the $\epsilon$ expansions at $\epsilon=1$ (and $n=3$) of the derivatives,  using  the procedure described in Appendix~\ref{app:resummation_strat}. Specifically, $\lambda^I_u\equiv\lambda^{}_u=a^{}_{10}$  and  $\lambda^I_v\equiv\lambda^{}_v=b^{}_{01}$ are the stability exponents of the isotropic FP, known accurately from resummations of the $\epsilon^6$ series~\cite{eps6}. Similarly, the isotropic exponents are identical to those calculated before, e.g. in Ref. \onlinecite{eps6}. The values in Table \ref{tab} are  taken from Refs. \onlinecite{KP17,KW20}, which use the same resummation as described in Appendix \ref{app:resummation_strat}.

\begin{table}[t]
 \centering
    \caption{Numerical estimates of  the coefficients entering Eqs. (\ref{RG1}), (\ref{RG2}) and (\ref{effexp}). The numbers are found by means of the resummation procedure described in Appendix~\ref{app:resummation_strat}. } %
    \label{one}
     \setlength{\tabcolsep}{9.1pt}
    \begin{tabular}{clcl}
      \hline
      \hline
      Quantity & Value& Quantity & Value  \\
      \hline
    $u^\ast_I$&$0.39273(63)$& $v^\ast_\textup{I}$& $0$         \\
    $a^{}_{01}$      &$-0.4791(19)$& $a^{}_{10}$      &$-0.7967(57)$\\
    $a^{}_{11}$      &$-0.938(36)$ & $a^{}_{02}$      &$-0.037(21)$ \\
    $a^{}_{20}$      &$-3.423(58)$ & $b^{}_{01}$      &$0.0083(15)$ \\
    $b^{}_{11}$      &$-1.854(15)$ &$b^{}_{02}$       &$-2.971(18)$ \\
    $\beta^{I}_{}$      &$0.3663(12)$\cite{KP17} &$\gamma^{I}_{}$ &$1.385(4)$\cite{KP17} \\
    $\varphi^{I}_{}$      &$1.263(13)$\cite{KW20}&       & \\
    $c^{}_{10}$      &$0.431(17)$ &$c^{}_{01}$       &$0.258(10)$ \\
    $c^{}_{11}$      &$0.49(11)$ &$c^{}_{20}$       &$0.82(20)$ \\
    $c^{}_{02}$      &$0.29(11)$ &$d^{}_{10}$       &$1.177(62)$ \\
    $d^{}_{01}$      &$0.706(38)$ &$d^{}_{11}$       &$0.820(54)$ \\
    $d^{}_{20}$      &$1.367(90)$ &$d^{}_{02}$       &$0.26(12)$ \\

    \hline
    \hline
    \end{tabular}
    \label{tab}
\end{table}

\begin{table}[t]
 \centering
    \caption{Numerical estimates of the cubic fixed point and  the  exponents obtained by our approximation, compared to those found in  
    Ref.~\onlinecite{eps6} and those calculated by means of the resummation strategy suggested in Ref.~\onlinecite{KP17} and used in Appendix \ref{app:resummation_strat}. The error bars on the approximate values are based on Eq. (\ref{effexp}).}
    \label{tab:vales_exponents}
     \setlength{\tabcolsep}{9.2pt}
    \begin{tabular}{llll}
      \hline
      \hline
      Quantity &Effective &Ref. \onlinecite{eps6}  &Using \\
      & & &Ref. \onlinecite{KP17}  \\
      \hline
    $u^*_{C}$ & $0.3791(27)$  & - & $0.376(19)$    \\
    $v^*_{C}$ & $0.0226(43)$  & - & $0.028(11)$     \\
    $\gamma^C_{}$ & $1.3849(61)$    &$1.368(12)$ & $1.387(9)$   \\
    $\beta^C_{}$  & $0.3663(21)$   &$0.3684(13)$ & $0.3669(12)$    \\
    \hline
    \hline
    \end{tabular}
    \label{tab1}
\end{table}

\begin{figure}[htb]
\centering
\includegraphics[width=.42\textwidth]{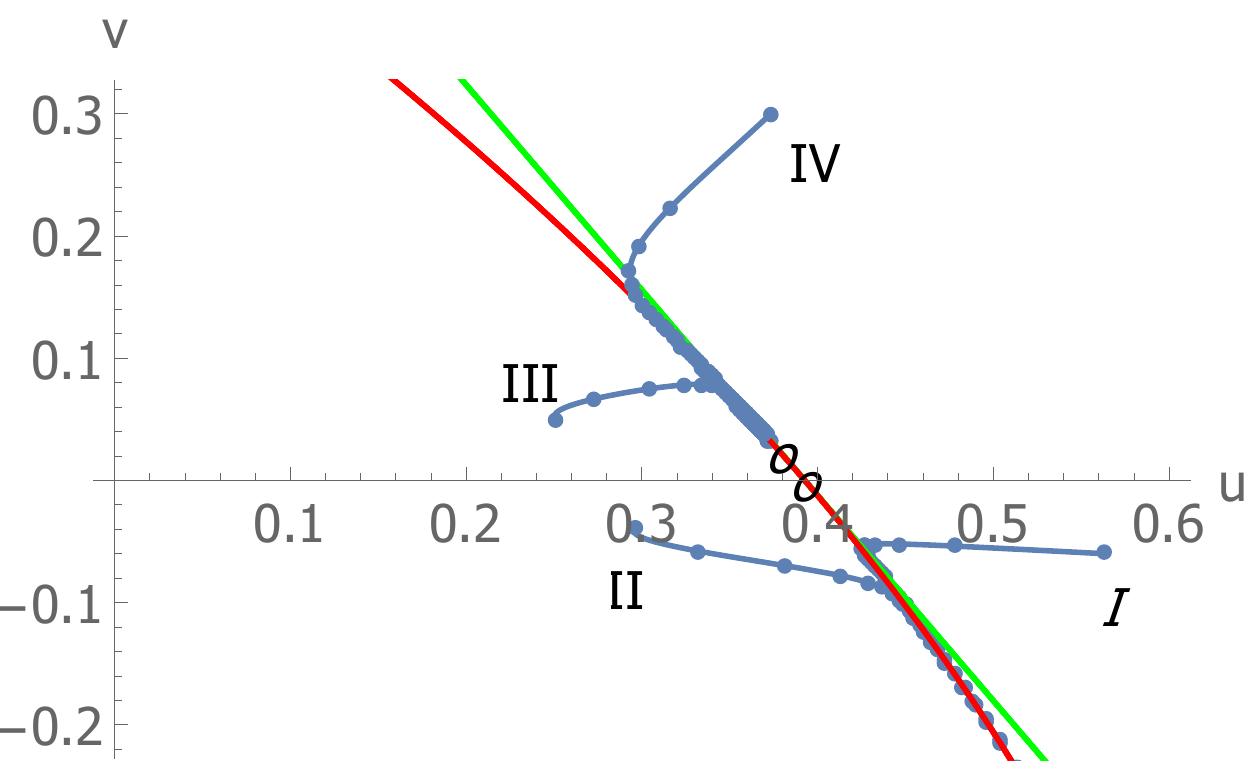}
\caption{(color online) Flow trajectories in the $u-v$ plane (blue) for several initial points. The dots indicate integer values of $\ell$. The red line is the  universal asymptotic line, Eq. (\ref{series}). The green line is the asymptotic line in the linear approximation, $v=-\delta u/z^{}_{01}$. The small circles denote the isotropic ($v^\ast_I=0$) and cubic fixed points. }
\label{f1}
\end{figure}

To solve  the recursion relations (\ref{RG1}) and (\ref{RG2}), it is convenient to define the non-linear scaling field~\cite{AF,comm},
\begin{align}
g^{}_u=\delta u +z^{}_{01} v+z^{}_{20} \delta u^2+z^{}_{11}
 v \delta u+z^{}_{02}v^2.
 \label{guu}
\end{align}
Here and below we keep only quadratic terms in $\delta u$ and $v$, since the recursion relations used that approximation. Using the coefficients
\begin{align}
&z^{}_{01}=a^{}_{01}/(\lambda^{}_u-\lambda^{}_v)\approx 0.595,\nonumber\\&z^{}_{11}=-\Big(a^{}_{11}-\frac{a^{}_{01} a^{}_{20}}{\lambda^{}_u}+\frac{a^{}_{01} b^{}_{11}}{\lambda^{}_u-\lambda^{}_v}\Big)\Big/\lambda^{}_v\approx -2.051\nonumber\\
&z^{}_{20}=-a^{}_{20}/(2\lambda^{}_u)\approx -2.148,\nonumber\\
&z^{}_{02}=\frac{a^{}_{02}/2+a^{}_{01}z^{}_{11}+b^{}_{02}z^{}_{01}/2}{\lambda^{}_u-2\lambda^{}_v}\approx -0.0983,
\end{align}
this scaling field obeys the linear equation
\begin{align}
\frac{\partial g^{}_u}{\partial\ell}=\lambda^{}_u g^{}_u,
\label{RGgu}
\end{align}
with the solution
\begin{align}
g^{}_u(\ell)=g^{}_u(0) e^{\lambda^{}_u \ell}.
\label{gul}
\end{align}
Since $\lambda^{}_u=a^{}_{10}=-0.7967(57)$, the exponential factor decreases fast with $\ell$.  Indeed, we show below that after a relatively small number of iterations $\ell^{}_1=\ln[\delta/g^{}_u(0)]/\lambda^{}_u$, with, say $g^{}_u(\ell^{}_1)=\delta=10^{-3}$, during which the RG  trajectories undergo a transient non-universal flow,  the trajectories approach a universal asymptotic line, on which they either flow to the cubic fixed point [$v(0)>0$] or to the first order region [$v(0)<0$]. Examples of  solutions are shown in Fig. \ref{f1}, where the asymptotic line is shown in red. This figure should replace the schematic Fig. \ref{fp} near the isotopic and cubic fixed points.

The explicit solution of the differential equations is presented in Appendix \ref{sol}. We first express $\delta u(\ell)$ in terms of $g^{}_u(\ell)$ and $v(\ell)$, from Eq. (\ref{guu}). Since $g^{}_u(\ell)$ quickly decays to zero, after a relatively small number of iterations $\ell^{}_1$, the flow quickly approaches an asymptotic line,
\begin{align}
\delta u(\ell)& \rightarrow -z^{}_{01}v(\ell)
-[z^{}_{02}+z^2_{01}z^{}_{20}-z^{}_{01}z^{}_{11}]v(\ell)^2\nonumber\\
&\approx -.595 v-.3613 v^2.
\label{series}
\end{align}
This line is {\it universal}, since its coefficients are fully determined by the isotropic fixed point.
Keeping terms up to quadratic order in $g^{}_u(\ell)$ and $v(\ell)$ yields a single differential equation, with the solution
\begin{widetext}
\begin{align}
\frac{1}{v(\ell)}=e^{-Ae^{\lambda^{}_u\ell}}e^{-\lambda^{}_v\ell}\Big[\frac{e^A}{v(0)}+
\frac{(-A)^{-\lambda^{}_v/\lambda^{}_u}B}{\lambda^{}_u}\Big(\Gamma[\lambda^{}_v/\lambda^{}_u,-A]
-\Gamma[\lambda^{}_v/\lambda^{}_u,-Ae^{\lambda^{}_u\ell}]\Big)\Big],
\label{solw}
\end{align}
\end{widetext}
where $A=b^{}_{11}g^{}_u(0)/\lambda^{}_u$, $B=-(b^{}_{11}/a+b^{}_{02}/2)$ and $\Gamma[s,z]$ is the  incomplete gamma function.

 Figure \ref{f1} shows the resulting RG trajectories, beginning at several initial points. We have checked that these analytic solutions coincide with a direct numerical solution of Eqs. (\ref{RG1})-(\ref{RG2}) for the range shown in the figure.
As seen, each RG trajectory has two (or three) main parts. In the first $\ell^{}_1$ iterations, $g^{}_u(\ell)$ decays quickly to zero, implying a fast  non-universal transient flow  towards the asymptotic line (\ref{series}). In this part, the points at integer values of $\ell$ are rather far from each other,  indicating the fast flow.  In the second part, the trajectory practically coincides with the asymptotic line. On this line, the points at integer values of $\ell$ become dense, indicating a slow variation with $\ell$. For $v>0$, this implies a slow approach to the cubic fixed point. For $v<0$, this  slow flow is followed by a third part, in which the flow gradually speeds up as the trajectory moves towards more negative values of $v$.

As explained in Appendix \ref{sol}, the flow at large $\ell$ can be described by
\begin{align}
v(\ell)=\frac{v(\ell^{}_1)e^{\lambda^{}_v(\ell-\ell^{}_1)}}{1+B v(\ell^{}_1)\big[e^{\lambda^{}_v(\ell-\ell^{}_1)}-1\big]/\lambda^{}_v}.
\label{vvv}
\end{align}
Since $\lambda^{}_v$ is very small, the variation of the second term in the denominator with $\ell$ is slow, explaining the second part above..
Since $B=0.3706>0$, this implies a slow variation in $v$ for positive $v$, approaching the cubic fixed point $v^\ast_C=\lambda^{}_v/B$. However, for $v(0)<0$ $v(\ell)$ becomes more negative. At first it decreases slowly, but when $e^{\lambda^{}_v \ell}$ becomes of order unity this decrease becomes faster (the points at integer $\ell$ become less dense),  and $v(\ell)$ diverges at $\ell=\ell^{}_2$, when $\big[e^{\lambda^{}_v(\ell^{}_2-\ell^{}_1)}-1\big]/\lambda^{}_v\approx \ell^{}_2-\ell^{}_1)=-1/[Bv(\ell^{}_1)]$. This value is larger for smaller $|v(\ell^{}_1)|$ [and therefore also for smaller $|v(0)|$].  Within our quadratic approximation, we are not allowed to follow this solution beyond some finite value, say $v<-.2$. However, it is reasonable that a full solution will also continue downwards, on the asymptotic trajectory, and reach the first order transition at $v(\ell)=-u(\ell)$.

\begin{figure}[htb]
\centering
\includegraphics[width=.4\textwidth]{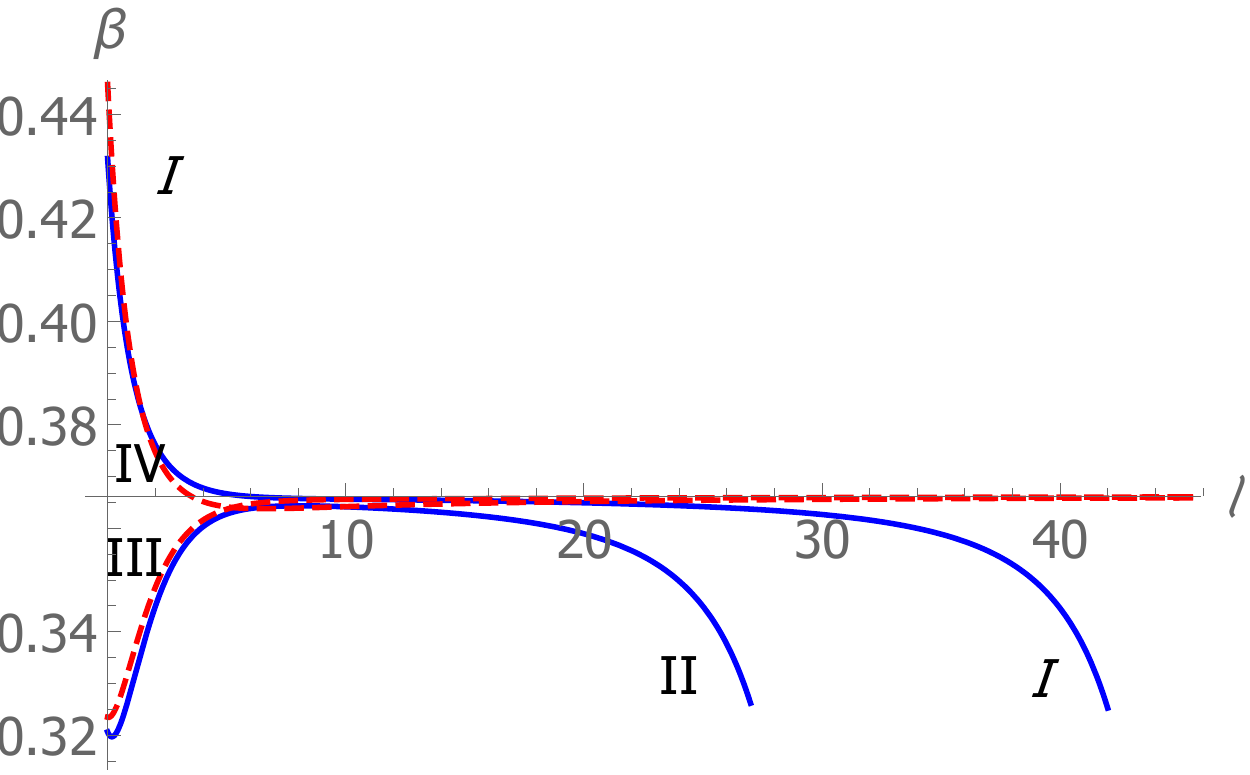}\ \ \includegraphics[width=.4\textwidth]{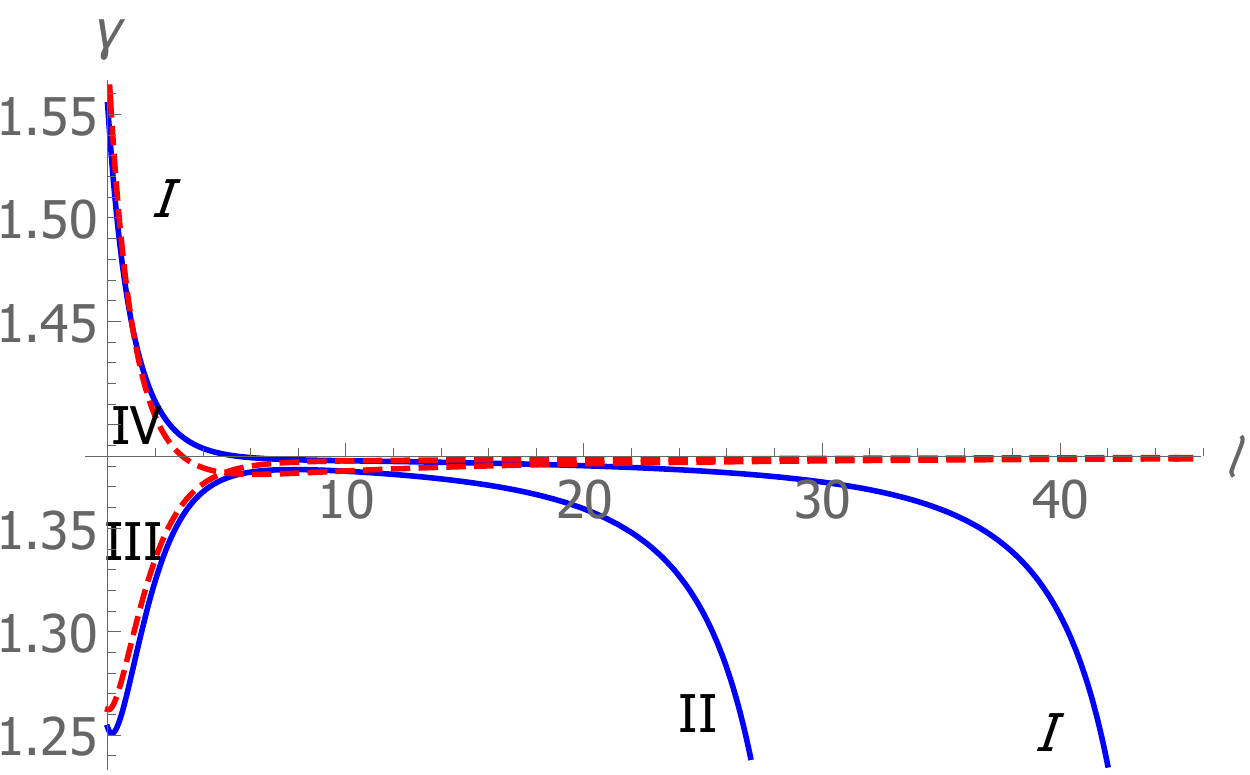}
\caption{(color online)  Effective exponents $\beta$ and $\gamma$ for the trajectories shown in Fig. \ref{f1}, as functions of $\ell$. The horizontal axes are at the asymptotic values of the isotropic fixed point (which are also very close to those at the cubic fixed point).  The exponents corresponding to trajectories with $v(0)>0$ (III and IV, dashed lines) approach this asymptotic value. In contrast, those with $v(0)<0$ (I and II, full lines)  initially come close to  these values, but then  turn downwards to smaller values.  }
\label{f2}
\end{figure}

The effective critical exponents $\beta$ and $\gamma$ are given by
\begin{align}
&\beta(u,v)=\beta^I+ c^{}_{10}\delta u+c^{}_{01}v+c^{}_{11}v\delta u
+\frac{c^{}_{20}}{2}\delta u^2+\frac{c^{}_{02}}{2}v^2,\nonumber\\
&\gamma(u,v)=\gamma^I+ d^{}_{10}\delta u+d^{}_{01}v+d^{}_{11}v\delta u
+\frac{d^{}_{20}}{2}\delta u^2+\frac{d^{}_{02}}{2}v^2,
\label{effexp}
\end{align}
with the coefficients listed in Table \ref{tab}. Figure \ref{f2} shows these effective exponents, calculated with $v(\ell)$ and $\delta u(\ell)$ from Eqs. (\ref{solw}) and (\ref{duell}). As expected, the exponents with $v(0)>0$ (dashed lines) approach the asymptotic values of the cubic fixed point (which are practically the same as those of the isotropic one). The rate of these approaches depends on the initial value $g^{}_u(0)$, and the corresponding effective exponents are smaller (larger) than the asymptotic ones if $g^{}_ u(0)<0~(>0)$. Both trajectories I and IV in Fig. \ref{f1} have $g^{}_u(0)>0$), and therefore both start at similar large effective exponents, and reach the vicinity of the isotropic (=cubic) fixed point values after a few iterations ($\ell^{}_1\sim 5$).   Generally, each value of $\ell$ is related to the initial value of $t$ via  $t(\ell)\sim t(0)e^{\ell/\nu}$. Assuming that the RG iterations end after $\ell^{}_f$ iterations, when $t(\ell^{}_f)\sim 1$, the asymptotic exponents can be observed  only if  $|t(0)|<e^{-\ell^{}_1/\nu}\sim e^{-5/\nu}\sim 10^{-3}$. Trajectories II and III start at similar negative values of $g^{}_u(0)$, and therefore they both  approach the asymptotic lines from below, resulting with effective exponents smaller than the asymptotic isotropic ones. After reaching the asymptotic line, the flow on that line is slow.
For trigonal systems, with $v(0)>0$, the effective exponents vary slowly on that line until they approach the asymptotic cubic values.

In the tetragonal case, $v(0)<0$, the beginning of the trajectories is similar to that described above, depending only on the sign of the initial $g^{}_u(0)$, which is not universal.  Once the asymptotic line is reached,  the initially slow growth of $|v|$, causes the effective exponents to approach slowly (mostly below) the isotropic asymptotic values, but then they turn downwards, diverging at a value  $\ell=\ell^{}_2$ that depends on the value of $v(\ell^{}_1)$.  Since the decrease in the exponents in part 3 of the flow is fully determined by the flow on the universal asymptotic line, and depends only on $v(\ell^{}_1)$, one can collapse the two full curves in Fig. \ref{f2} onto each other just by shifting them along the $\ell$ axis. In particular, the decreasing part of line of curve I accurately overlaps that of line II when we shift $\ell \rightarrow \ell-15$. This reflects a universality of these effective exponents. We are not aware of earlier discussions of such a universality.

 We stopped our calculation when the trajectory left the region $|v|<0.2$, where our quadratic approximation may fail.  However, we expect that the asymptotic  universal line will eventually cross the line $v=-u$, and the transition will become first order. Given Eq. (\ref{vvv}), this will happen at larger $\ell$, i.e., smaller $t$, when $|v(0)|$ is smaller. Since SrTiO$^{}_3$ is supposed to have a very small negative $v$ (see below), this may explain why its first order transition happens at a very small $t$, which has not yet been reached experimentally.
In contrast, KMnF$^{}_3$ and  RbCaF$^{}_3$ do show first order transitions at some finite $t$, implying that they start at larger values of $|v(0)|$.

\section{Experiments}

Based on the mean-field region of the experiments, M\"{u}ller {\it et al.}~\cite{KAM-NATO} estimated the `initial' Landau parameters  to be $\{u,v\}^{}_{\rm S}\cong \{1.91, -0.068\}$ for SrTiO$^{}_3$ and  $\{u,v\}^{}_{\rm L}\cong 0.06\pm 0.06,~ 0.68\pm 0.06\}$ for LaAlO$^{}_3$, all in cgs units divided by $10^{43}$. These rough values (which should be improved!) fit beautifully with our expectations: SrTiO${}_3$ (S in Fig. 2) has a small initial negative $v$, and we predict that it will  flow  quickly parallel to the horizontal axis towards the universal asymptotic line and the isotropic FP before turning downwards, as for trajectory I in Fig. \ref{f1}. 

On the other hand,  LaAlO$^{}_3$ (L in Fig. 2) starts close to the positive $v-$axis, in a region which is probably not covered by our quadratic approximation.
In principle, one could perform an expansion of the  RG recursion relations near the decoupled Ising fixed point. Since the instability exponent at that fixed point, $\lambda^D_u=\alpha^D/\nu^D$ (with the Ising values of $\alpha$ and $\nu$~\cite{DG}) is small, we again expect two parts for the flows which begin close to the $v$-axis. At first, the trajectories will flow quickly to the vicinity of the decoupled fixed point, dominated by the large negative exponent $\lambda^D_v$(=the stability exponent of the Ising FP), approaching a universal asymptotic line on which they will flow slowly away from the decoupled FP. Eventually, this asymptotic line will coincide with the one we calculated near the isotropic FP, and the flows will approach the cubic fixed point along that line, as  in trajectories III and IV in Fig. \ref{f2}. Interestingly, the line (\ref{series}) reached the $v-$axis at $v\approx 0.5$, quite close to the decoupled FP value  $v^\ast_D=0.482(9)$ which we calculated for $n=1$ using the resummation technique of Appendix \ref{app:resummation_strat}. Therefore, Eq. (\ref{series}) may give a good approximation for the whole universal line connecting the D, C and I fixed points.

To the best of our knowledge, there are no experiments showing first-order transitions into the trigonal phase~\cite{LAO}, but -- as listed in Sec. \ref{Intro} -- several experiments  found such transitions into the tetragonal phase. Interestingly, the experiments listed in Sec. \ref{Intro}
 exhibit intermediate regions (before the first-order transitions) with effective critical exponents, which are smaller than their isotropic values. This is consistent with our Fig. \ref{f2}.  The  smaller values of $\beta$ in NaNbO$^{}_3$ have also been attributed to the inter-plane weak correlations along the rotation axis~\cite{Lifshitz}.

 To test our predictions in detail, we propose a repeated analysis of existing experiments, and new dedicated experiments, in which the critical exponents will be measured over several separate ranges of $t$, to check how they vary with $t$ (which is equivalent to their dependence on the number of iterations  $\ell$).

As mentioned in Sec. \ref{Intro}, polishing the surfaces of the crystals SrTiO$^{}_3$ and LaAlO$^{}_3$ caused a crossover from the $n=3$ critical behavior to that of the Ising ($n=1$) critical behavior. This crossover is due to an axially anisotropic Hamiltonian, ${\cal H}^{\alpha\beta}_{g}=g\big(Q^{}_\alpha Q^{}_\beta-\delta^{}_{\alpha\beta}{\bf Q}^2/3\big)$, where $g$ represents the uniaxial stress. Such uniaxial stress has also been applied directly, yielding information on the  corresponding exponent $\lambda^{}_g$.
Existing experiments gave a range of values for $\lambda^{}_g$. A detailed application of our theory to these exponents  will be presented separately.

 One way to vary $v$ experimentally is to use mixed crystals, e.g. Sr$^{}_{1-x}$Ca$^{}_x$TiO$^{}_3$~\cite{mixed}  or a mixture of SrTiO$^{}_3$ with LaAlO$^{}_3$, which is expected to be easy to grow due to their matching lattice constants~\cite{STO-LAO}.  Since  both the isotropic and cubic FP's have $d\nu>2$, randomness is irrelevant~\cite{ABH,rAA} and one expects the same competition predicted above. %\st{The properties of many mixed
It is interesting to note that KMn$^{}_{1-x}$Ca$^{}_x$F$^{}_3$ seems to approach a second-order transition as $x$  increases~\cite{KTCaO}. If the transition is still into the tetragonal structure, this may represent a smaller value of the initial $|v|$  in the dilute case. This may be caused by the larger dimension of the parameter space in the dilute case, which involves many transient iterations until the flow reaches the $u-v$ plane ~\cite{rAA}.

Three-component order-parameter vectors ${\bf Q}$ with cubic symmetry are  abundant. Examples include  ferroelectric transitions from cubic to tetragonal or trigonal,  ferromagnets and antiferromagnets with the magnetization ordering along an axis or a diagonal. Close to their transitions, these cubic crystals may be described  by Eq. (\ref{H0}), and therefore their critical behavior is expected to split into the two types described above.  In particular, {\it ordering of the $n=3$ cubic system along a cube axis must be fluctuation-driven first-order, with intermediate effective isotropic exponents!} This important prediction can be tested experimentally~\cite{com-vic}.

\section{Conclusions}

In conclusion, we have shown that for systems having cubic symmetry, universality is not determined merely by the  symmetry above the transition. Rather, the critical behavior of such systems depends on their symmetry {\it below} the transition. Cubic systems split into two groups - those that become trigonal, which belong to the cubic FP universality class, and those that become tetragonal, which undergo a fluctuation driven first-order transition. In both cases, there is a wide temperature range in which slow-varying  effective exponents, which exhibit universal features,  are expected.  For systems like SrTiO$^{}_3$, the cubic to tetragonal transition exhibits effective exponents close to those of the isotropic FP, but then cross over to a fluctuation-driven first-order transition. This crossover is accompanied by large changes in the effective exponents, which await experimental detection.
These results  resolve a long standing confusion about the universality of the displacive phase transitions in the perovskites, and leaves its complete confirmation to future dedicated experiments, concentrating on the $t-$ dependence of the effective exponents.

\begin{acknowledgments}
This research was initiated by a very stimulating discussion with Slava Rychkov, who drew our attention to the new accurate values of $n^{}_c$ (see also Ref. \onlinecite{paris}). A.K. gratefully acknowledges Mikhail Kompaniets for the helpful discussion and Andrey Pikelner for his help with RG expansions from Ref.~\onlinecite{below}, and
 the support of Foundation for the Advancement of Theoretical Physics "BASIS" through Grant 18-1-2-43-1.

\end{acknowledgments}

%%%%%%%%%%%%%%%%%%%%%%%%%%%%%%%%%%%%%%%%%%%%%%%%%%%%%%%%%%%%55
%%%%%%%%%%%%%%%%%%%%%%%%%%%%%%%%%%%%%%%%%%%%%%%%%%%%%%%%%%%%55

\appendix

\section{Details of the resummation}\label{app:resummation_strat}

Since the $\epsilon-$expansions, e.g. $f(\epsilon)=\sum_{k=0}^\infty f^{}_k\epsilon^k$, are divergent, numerical estimates of the quantities of interest at $d=3$ are obtained employing proper resummation techniques~\cite{brezin1,LIPATOV}. Such resummations were performed for the isotropic and cubic FP's in  Ref.~\onlinecite{eps6},  which used   the basic resummation procedure of the Pad\'e-Borel-Leroy method. Here we use the $\epsilon-$expansions based on those in Ref.~\onlinecite{eps6}, but apply the more advanced  resummation strategy proposed in Ref.~\onlinecite{KP17}.
 The main idea of this strategy can be formulated simply: the asymptotic behavior of the series coefficients for large order is written as
 \begin{align}
 \label{eqn:hoab_A_expansion}
  f_k \xrightarrow[k \rightarrow \infty]{} c\, k! k^{b^{}_0}(-a)^k ,
\end{align}
where $1/a$ is the radius of convergence and $b^{}_0$ is fixed by the high-order asymptotic behavior of the series. However, in practice we only have a limited number of terms in the series,  and consequently  this asymptotic behavior is not known. Therefore, the authors of Ref. \onlinecite{KP17} proposed to treat the Leroy parameter $b=b^{}_0+3/2$ as a free parameter, to be determined variationally (see below).

 Once a Borel transformation, based on this modified asymptotic form, is performed, the variable $\epsilon$ is conformally mapped onto
\begin{align}
w(\epsilon)=\dfrac{\sqrt{1+a\epsilon}-1}{\sqrt{1+a\epsilon}+1}, \quad \epsilon(w)=\dfrac{4w}{a(1-w)^2},
\end{align}
and it is assumed that the function has the strong asymptotic behavior $f(\epsilon)\sim \epsilon^\lambda,\ \ \epsilon\rightarrow \infty$.
The results are then improved by a preliminary homogeneous homographic transformation,
\begin{align}
\label{eqn:homographic_trans}
\epsilon(\epsilon')  \rightarrow \dfrac{\epsilon'}{1+q\epsilon'}, \quad \epsilon'(\epsilon)  \rightarrow \dfrac{\epsilon}{1-q\epsilon},
\end{align}
and the final approximate estimates are found by applying the steps mentioned above to the new $\epsilon'$ expansion.
Finally,
 the optimal values of the parameters $b,~\lambda$ and $q$  (for each series) are determined by the least sensitivity to their variation (for details see Ref.~\onlinecite{KP17}). This technique was applied to find all the coefficients listed in Table \ref{tab}.

\section{Solution of the differential equations}\label{sol}

Solving Eq. (\ref{guu}) for $\delta u(\ell)$ yields
\begin{widetext}
\begin{align}
 du(\ell)=\Big[-1-z^{}_{11} v(\ell)+\sqrt{[1+z^{}_{11}v(\ell)]^2-4z^{}_{20}[z^{}_{01}v(\ell)+z^{}_{02}v(\ell)^2-g^{}_u(\ell)]}\Big]\Big/(2z^{}_{20}).
 \label{dul}
\end{align}
Expanding to quadratic order in $v(\ell)$ and $g^{}_u(\ell)$ gives
\begin{align}
\delta u(\ell)\approx g^{}_u(\ell)-z^{}_{20}g^{}_u(\ell)^2-\big[z^{}_{01}+\big(z^{}_{11}-2z^{}_{01}z^{}_{20}\big)g^{}_u(\ell)\big]v(\ell)
-\big[z^{}_{02}+z^{2}_{01}z^{}_{20}-z^{}_{01}z^{}_{11}\big]v(\ell)^2.%\nonumber\\
%&=g^{}_u(\ell)+\frac{a^{}_{20}}{2\lambda^I_u}g^{}_u(\ell)^2
%-\big[\frac{a^{}_{01}}{\lambda^I_u}+\big(z^{}_{11}+\frac{a^{}_{01}a^{}_{20}}{(\lambda^I_u)^2}\big)g^{}_u(\ell)\big]v(\ell)
%+\frac{a_{01}^2a^{}_{20}- a^{}_{01}b^{}_{02}\lambda^I_u-a^{}_{02}(\lambda^I_u)^2}{2(\lambda^I_u)^3}v(\ell)^2.
\label{duell}
\end{align}
\end{widetext}
After a few iterations $g^{}_u(\ell)$ decays to zero, and this solution reaches the asymptotic line (\ref{series}).

We now return to Eq. (\ref{RG2}). Substituting Eq. (\ref{duell}), and stopping at quadratic order, this equation becomes
\begin{align}
\frac{\partial v}{\partial\ell}&\approx \lambda^{}_v v(\ell)+b^{}_{11}\big[g^{}_u(\ell)-z^{}_{01}v(\ell)\big]v(\ell)
+b^{}_{02}v(\ell)^2/2\nonumber\\&=\lambda^{}_v v(\ell)+b^{}_{11}g^{}_u(\ell)v(\ell)-B v(\ell)^2,
\label{eqv}
\end{align}
where $B\equiv b^{}_{11}z^{}_{01}-b^{}_{02}/2$. This can be written as
\begin{align}
\frac{\partial}{\partial\ell}\Big[\frac{1}{v}\Big]=-\big[\lambda^{}_v+b^{}_{11}g^{}_u(0)e^{\lambda^{}_u\ell}\big]\Big[\frac{1}{v}\Big]+B.
\label{solv}
\end{align}
To solve this equation, we write $x=e^{\lambda^{}_u \ell}$ and
\begin{align}
\frac{1}{v(x)}=e^{-A x} x^{-\lambda^{}_v/\lambda^{}_u}W(x),
\end{align}
where $A=b^{}_{11}g^{}_u(0)/\lambda^{}_u$. This yields
\begin{align}
\frac{dW}{dx}=\frac{B}{\lambda^{}_u}e^{Ax}x^{\lambda^{}_v/\lambda^{}_u-1},
\end{align}
with the solution (\ref{solw}), where
\begin{align}
\Gamma[s,z]=\int_z^\infty t^{s-1}e^{-t}dt.
\end{align}
is the incomplete gamma function.

For large $\ell$, $x=e^{\lambda^{}_u\ell}$
is small, and  $\Gamma[s,z]=-z^s/s+{\cal O}[1]$, so that $\Gamma[\lambda^{}_v/\lambda^{}_u,-Ax]\propto e^{\lambda^{}_v\ell}$.  This result can be obtained directly:   For $\ell>\ell^{}_1$  we can neglect $g^{}_u(\ell)$  in Eq. (\ref{solv}). The solution to this equation is then given in Eq. (\ref{vvv}).

%%%%%%%%%%%%%%%%%%%%%%%%%%%%%%%%%%%%%%%%%%%%%%%%%%%%%%%%%%%
%%%%%%%%%%%%%%%%%%%%%%%%%%%%%%%%%%%%%%%%%%%%%%%%%%%%%%%%%%%


\begin{thebibliography}{99}
\bibitem{com-vic} The distinction between ordering along an axis or along a diagonal was already noticed in Ref. \onlinecite{vic-rev}. Here we emphasize the variation of the effective exponents for both cases, and the large variations of these exponents in the latter fluctuation-driven first-order case.
\end{thebibliography}
\end{document}